# CHANNELING OF HIGH ENERGY BEAMS IN NANOTUBES

S. Bellucci[1], V.M. Biryukov[2], Yu.A. Chesnokov[2], V. Guidi[3], W. Scandale[4]

[1]INFN - Laboratori Nazionali di Frascati, P.O. Box 13, 00044 Frascati, Italy

[2]Institute for High Energy Physics, Protvino, Ru-142281, Russia

[3]Department of Physics and INFN, Via Paradiso 12, I-44100 Ferrara, Italy

[4]CERN, Geneva 23, CH-1211, Switzerland

Presented at **COSIRES 2002** (Dresden, 24-27 June 2002)

**Corresponding author:** V.M. Biryukov; E-mail: biryukov@mx.ihep.su

## Abstract

We present simulations of particle beam channeling in carbon nanotubes and evaluate the possibilities for experimental observation of channeling effect in straight and bent nanotubes at IHEP and LNF. Different particle species are considered: protons of 1.3 and 70 GeV, and positrons of 0.5 GeV. Predictions are made for the experiments, with analysis of requirements on the quality of nanosamples and resolution of the experimental set-up. Based on Monte Carlo simulations, the capabilities of nanotube channeling technique for particle beam steering are discussed.

## Introduction

Channeling of particle beams in crystal lattices finds many applications in accelerator world

[1] from TeV [2] down to MeV [3] energies. In accelerator ring, bent crystals serve for beam deflection, e.g. in extraction systems in IHEP-Protvino [4] and in collimation system of Relativistic Heavy Ion Collider of BNL [5]. This technique can be quite efficient. In IHEP, a tiny 2-mm Si crystal did extract 70-GeV protons out of the ring with efficiency 85%, even when all the beam stored in the machine was dumped onto the crystal, at intensity well over $10^{12}$ p/s [6].

For channeling, silicon crystals are commonly used due to high perfection of their lattice and availability. Germanium crystals were also demonstrated as efficient deflectors of high-energy beams [7]. There is an interest in using also low-Z and high-Z crystals for beam steering [8]; bent diamond crystals have been produced [9], and there is an effort to produce tungsten crystals of a quality sufficient for beam steering applications [10].

In recent years, with creation of nanotubes, there has been a substantial interest in channeling in nanostructures[11-13]. Since 1991 [14], there has been a lot of study on carbon nanotubes to understand their formation and properties. Carbon nanotubes stick out in the field of nanostructures, owing to their exceptional mechanical, capillarity, electronic transport and superconducting properties [15-17]. They are cylindrical molecules with a diameter of order 1 nm and a length of many microns [18]. They are made of carbon atoms and can be thought of as a graphene sheet rolled around a cylinder [19]. Nanotubes can be made of different diameter, length, and even material other than carbon [20]. Creation of suitable channeling structures, from single crystals to nanotubes, of sufficient quality would make a strong impact onto accelerator world.

A particle beam channeled in nanotube could be efficiently steered (deflected, focused, undulated, extracted from accelerator, etc.) in the way quite similar to crystal channeling. Compared to bent crystal technique, however, nanotubes offer unique characteristics like bigger angular acceptance and potentially bigger dechanneling lengths. This may be used to create a very elegant technique of beam handling at accelerators. All advantages of the crystal technique application at accelerators, like low cost, low size, minimal "septum width" of crystal deflectors, remain the same also for nanotube technique. Besides their usage as beam deflectors in extraction or collimation systems (potentially at Large Hadron Collider or at medical accelerators), nanotubes can boost new applications like very small nano-beams for biological studies and medical therapy. This would depend strongly on the capabilities of nanotechnology to produce required structures for beam steering.

## Nanotube channeling

The feasibility of channeling in nanotubes has been earlier proposed, e.g., in refs. [11-13]. It was shown [13] that a bent carbon nanotube of 1.4 nm diameter has a significant effective potential well $U_{eff}$ even for bendings equivalent to $\cong$ 300 Tesla or $pv/R \geq$ 1 GeV/cm (the beam momentum ratio to curvature radius), so there exists an opportunity to steer particle beams by nanotubes, similarly to bent crystal channeling technique.

We have performed computer simulations [21] of nanotube channeling to evaluate the achievable performance of a nanotube as a steering device. In simulations, we used so-called standard potential introduced by Lindhard [22]. When averaged over the longitudinal coordinate and azimuth angle ($\varphi$, z), the potential of nanotube is described by [11]:

$$U(\rho) = \frac{4NZ_1Z_2e^2}{3a} \ln\left( \frac{\tilde{r}^2 + \tilde{\rho}^2 + \tilde{\varepsilon}^2 + \sqrt{(\tilde{r}^2 + \tilde{\rho}^2 + \tilde{\varepsilon}^2)^2 - 4\tilde{r}^2\tilde{\rho}^2}}{2\tilde{r}^2} \right)$$

Here $Z_1e$, $Z_2e$ are the charges of the incident particle and the nanotube nuclei respectively, $N$

is the number of elementary periods along the tube perimeter, $a=0.142$ nm is the carbon bond length; dimensionless parameters are

$$\tilde{r} = \frac{r}{3a}, \quad \tilde{\rho} = \frac{\rho}{3a}, \quad \tilde{\varepsilon} = \frac{a_s}{a\sqrt{3}}$$

the nanotube radius is

$$r = \frac{Na\sqrt{3}}{2\pi}$$

The screening distance $a_S$ is related to the Bohr radius $a_B$ by

$$a_s = \frac{a_B}{2} \left( \frac{3\pi}{4(\sqrt{Z_1}+\sqrt{Z_2})} \right)^{\frac{2}{3}}$$

In a tube bent along the $x$ direction, the motion of a particle is described by the equations

$$pv\frac{d^2x}{dz^2} + \frac{dU(\rho)}{dx} + \frac{pv}{R(z)} = 0$$

$$pv\frac{d^2y}{dz^2} + \frac{dU(\rho)}{dy} = 0$$

where $\rho^2 = x^2 + y^2$. This takes into account only the nanotube potential and the centrifugal potential. Any particle within close distance, order of $a_S$, from the wall (where density of the nuclei is significant) is also strongly affected by the nuclear scattering.

In our simulations, we firstly were interested in the persistence of channeling in *bent* nanotubes, for two major reasons: (a) we need nanotubes to steer and bend particle beams, and (b) real nanotubes are never exactly straight. Figure 1 shows the results of our simulations for the number of channeled protons as a function of the nanotube curvature *pv/R* for tubes of different diameter (0.55, 1.1, and 11 nm). For comparison, also shown is the same function for Si(110) crystal. These results are obtained for a parallel incident beam and a uniformly bent channel (without straight part).

One can see from Figure 1 that nanotubes should be sufficiently narrow in order to steer efficiently the particle beams; the preferred diameter is in the order of 0.5-2 nm. Wider nanotubes, like 10-50 nm, appear rather useless for channeling purpose because of their high sensitivity to channel curvature. When compared to bent crystals as elements of beam steering technique, bent nanotubes show in simulations good efficiency of beam bending, similar to that of crystals. The Monte Carlo code used here for simulation of nanotube channeling has been adapted from the CATCH code [23], and is described in some detail elsewhere [21].

## Optimisation of the experiment

The purpose of the present work is to propose an experiment for observation of nanotube channeling of particle beam and to find an experimental arrangement most suitable for observation of channeling in straight and bent nanotubes. As the typical nanotubes are limited in the length at present (to order of 50 μ m), this limits their use at higher energies like 70 GeV where most of the IHEP experimental research in crystal channeling has been done so far. Therefore, two approaches to experimental verification of the channeling effect are considered.

The first one is based on the observation of coherent scattering of particles off the potential of straight nanotubes aligned to the incident beam. If we minimize any extra material in the beam such as substrate, the coherent scattering appears much stronger than the usual multiple scattering and it should be seen only within the angular range of about the critical channeling angle as we rotate the nano-sample (plus the divergence of the nanotube alignment at the sample entrance).

The second approach assumes that nanotubes trap and channel part of the incident beam. By giving to nanotubes a controlled bending of a few milliradian, we can deflect the channeled particles out of the incident beam (same as in bent crystal channeling). The deflected beam could be easily observed; like in the previous case, it shows up only for aligned sample. This approach seems easier for observation, however it requires more skills in engineering and production of the sample.

Accordingly, two sorts of sample have to be designed, simulated, produced, and tested.

## Straight nanotubes

In simulations, we take what we believe is our best for the parameters of the beam and the experimental set-up. The full width of the incident beam of 70 GeV protons is taken 0.1 mm (one microstrip), the divergence ±0.040 mrad full width (0.023 mrad rms) both horizontally and vertically. After interaction with a nano-sample, the particles are to be detected by a microstrip detector (0.1 mm strips) positioned 8 m downstream. Most of this 8 m is vacuum; the Mylar windows and some air gaps are equivalent in total to 1 m of air. The multiple scattering on this 1 m of air is taken into account in simulations, as well as the multiple scattering on C atoms in nano-sample.

Figure 2 shows an example of the beam distribution as observed at the microstrip detector when the sample is misaligned to the beam; the starting distribution is just widened by multiple scattering and by beam divergence. To observe the effects of coherent scattering off the potential of nanotubes with sufficient clarity, we must carefully choose the best length of the nanotubes. The angular kick as provided by the nanotube field to the particles depends on the particle entrance coordinate with respect to the tube axis, but also depends on the distance traveled along the tube. This distance is somewhat related to the oscillation wavelength of a particle channeled in the nanotube. Obviously, the largest kick would be if the tube were as long as about ¼ of a wavelength. Unlike in crystal channeling where planar potentials are close to harmonic, in nanotube channeling the oscillation wavelengths are quite dependent on the particle starting point. We have done the optimisation for the length of nanotubes (with the tube diameter fixed at 1.1 nm) in Monte Carlo simulations.

Figure 2 shows also an example of the beam distribution at the microstrip detector when the sample of 30-•m long nanotubes is perfectly aligned to the above-defined 70-GeV beam. Notably, the distribution becomes factor of three broader than with misaligned sample, giving a clear signature of the coherent scattering off the nanotube potential. When the sample is slightly tilted with respect to the beam, the distribution remains broad but also obtains a clear asymmetry as illustrated by Figure 3 for the same sample angled at 0.040 mrad with respect to the beam. This asymmetry, depending on the sign and magnitude of the tilt, is another good signature for the coherent effects in nanotubes.

The dependence of the distribution broadening and asymmetry on the nanotube length is shown in Fig.4. Precisely, the left part of Fig.4 shows the rms size of the distribution for aligned sample, normalised to the same quantity for misaligned sample. The right part in Fig.4 shows the change in the mean value of the distribution when sample is angled at 0.040 mrad to the beam.

For the aligned sample, the broadening of the distribution is the largest for the nanotubes 25-50 micron long; outside of this range, it is less pronounced. For the sample aligned at 0.040 mrad with respect to the beam, the particles are best scattered with nanotubes of 20-50 micron length; the asymmetry is the largest for nanotubes of 25-40 micron. The coherent effects are clearly seen within the angular range of about 0.150 mrad (full width); outside of this range of the nanosample orientation, there is little to see.

To summarise, we find the best length to be 30 micron in our set-up. The amorphous substance in the sample is not so dangerous. The multiple scattering over even 100 micron of carbon is yet small at 70 GeV. On the other hand, the angular divergence of the tubes is important. If the tubes in the sample are diverging over much more than 0.150 mrad, we shall see significantly lesser signal. The solution might be to have nanotubes in a *corset*, inside a porous material. The amount of the substrate or holder is not so important, but the divergence is important at 70 GeV. The tubes can be capped by substrate on both ends if necessary to ensure their parallelism. Another experimental factor is the density of the tubes in the sample. In the above simulation the tubes cross-section was 78% of the total (the densest pack would make it 91%). Then, about 1/2 of the beam was "channeled", i.e. found outside of the central three strips of the detector. If the density of the tubes would be 10 times smaller (i.e. 8% of the sample cross-section), we would observe only 5% of the beam out of the center 3 strips, accordingly. Let us take this consideration in a more general way: we are interested in the density of parallel nanotubes; one can have more tubes in the sample - but if they are misaligned, they do not contribute.

## Bent nanotubes

To bend a channeled beam with a nanotube, we should use a curvature no greater than $pv/R \leq 1$ GeV/cm. This means using either long enough nanotubes or lower energy of the particles. In the experiment, we will be limited by the nanotube length of about 60 micron. Therefore, we consider experiments with protons of 1.3 GeV kinetic energy and positrons of 0.5 GeV. Then, 60 micron is enough to give particles a deflection of a few milliradian.

The experiment with 1.3 GeV protons might have been done with the beam circulating in the main ring of U70 accelerator. The incident particles would have then quite low divergence at the incidence on the nanotube deflector. We have done simulations with the particles incident at different angles at a 60-•m long nanotube bent 2 mrad, for 1.3 GeV protons and 0.5 GeV positrons, observing the angular distribution of particles downstream of the nanotube. If the incident particles are well aligned to the nanotube axis at the entrance, a substantial amount of particles is deflected the full angle of bending. Figure 5 shows the simulated distribution of 1.3-GeV protons at the detector (a hodoscope with 1.25-mm strips as used in the previous channeling experiments [6] with 1.3 GeV protons) placed 20 m downstream of the bent nanotube. The peak seen is the channeled protons; the background from the multiply scattered nonchanneled protons circulating in the ring, expected to appear at the edge of the detector on the most-left strips, is not fully simulated here. An experimentally observed distribution in the same set-up but with a silicon crystal can be seen in Ref. [6].

In order to understand the requirements for orientation match between the nanotubes and the incident particles, we varied the tube orientation in simulations and present the results in

Figure 6. The channeling effects are seen within the angular range of about 0.6 mrad (full width) for 1.3 GeV protons and about 0.8 mrad for 0.5 GeV positrons. This limits the allowed divergence of particles in the incident beam or of the nanotubes within the sample to the above-said level in order to have a sizeable effect in the experiment.

## Summary

As shown in computer simulations, nanotubes can efficiently channel particle beams thus making a basis for a new technique of beam steering at accelerators. The most critical factor in implementation of this technique is the alignment of nanotubes within the sample, and to some extent the divergence of the particle beam. From physics standpoint, this is an excellent technique, however it is challenging technologically. It is of high interest for channeling research as it looks that nanostructures can be engineered to fit researcher the most, with wide choice of material, lattice design, etc.

## Acknowledgements

This work was partially supported by INFN - Gruppo V, as NANO experiment, and by INTAS-CERN Grant No. 132-2000 and RFBR Grant No. 01-02-16229.

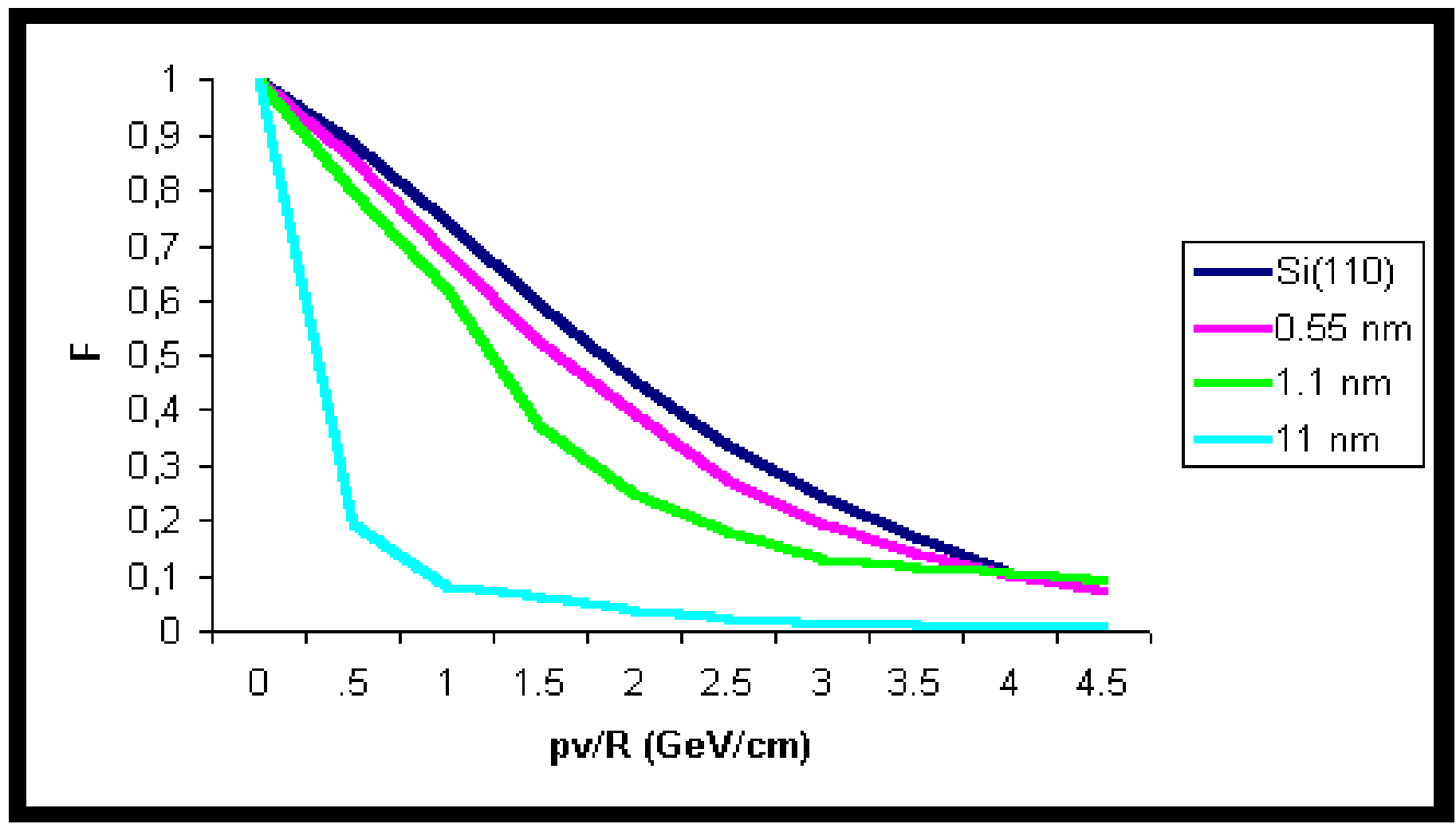

**Figure 1** The number of channeled protons shown as a function of the nanotube curvature *pv/R* for tubes of different diameter (0.55, 1.1, and 11 nm). For comparison, also shown is the same function for Si(110) crystal.

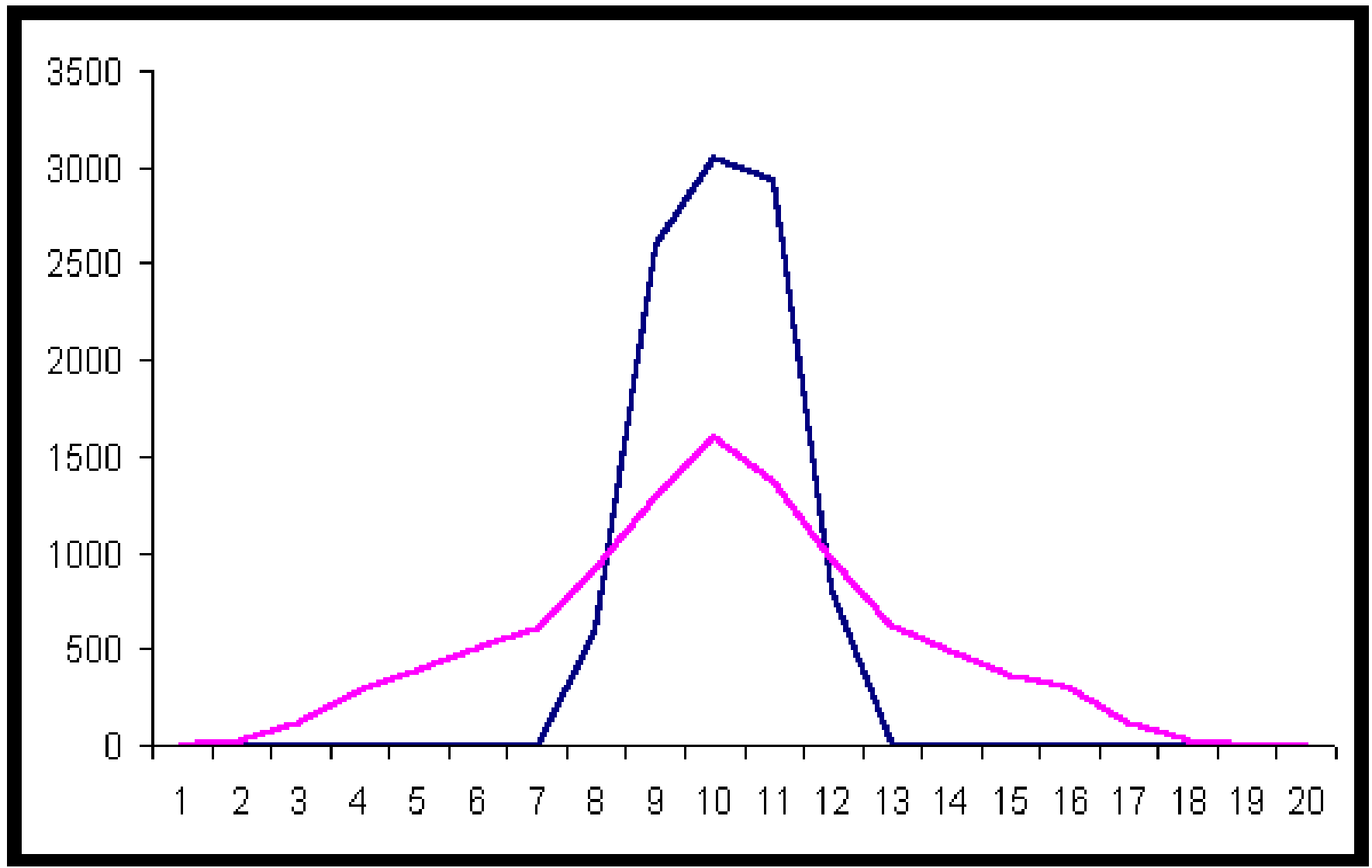

**Figure 2** Beam distribution at the microstrip detector for the misaligned sample (narrow peak), and for the sample perfectly aligned to the beam (broad peak).

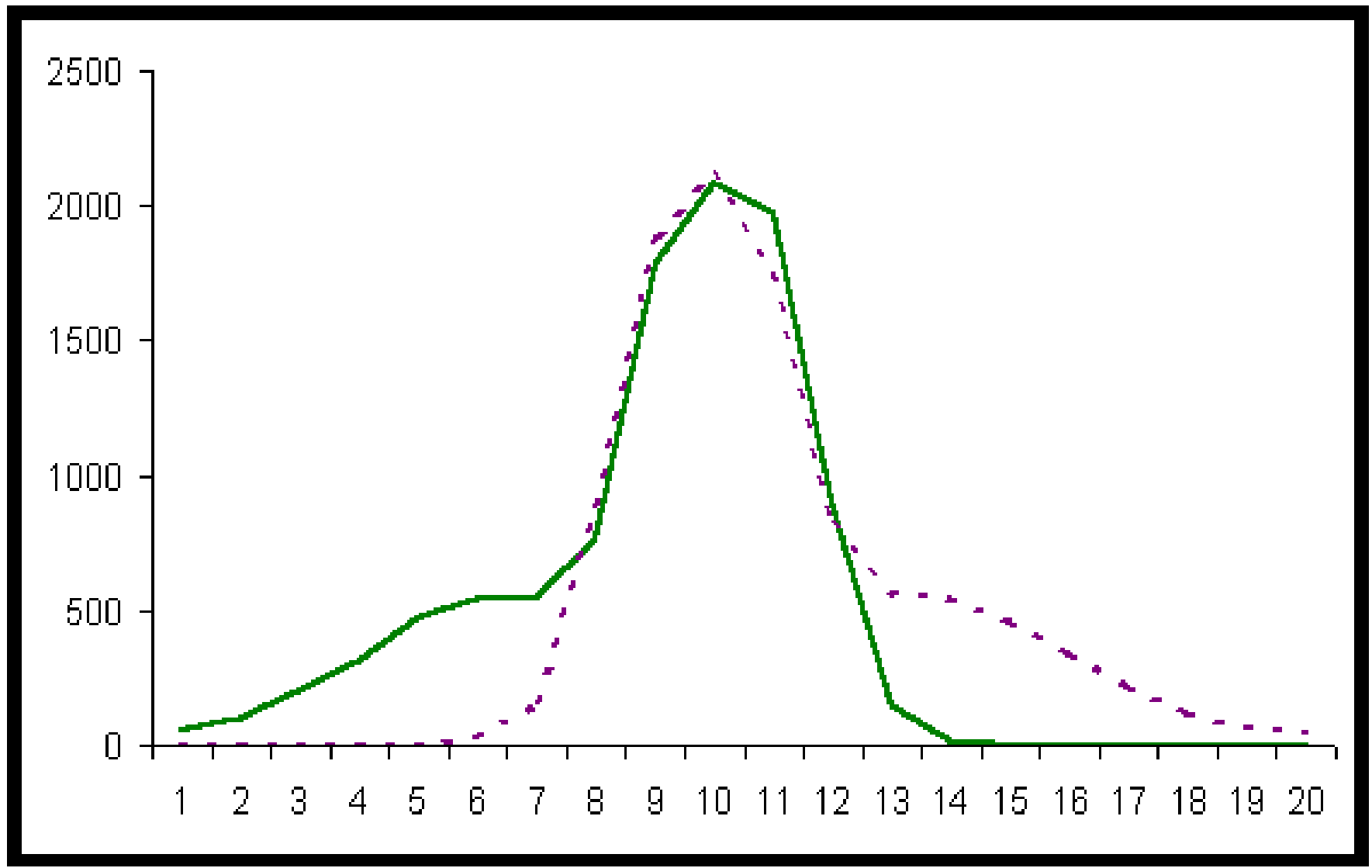

**Figure 3** Beam distribution at the microstrip detector for the sample aligned at +0.04 mrad to the beam (solid line), and at -0.04 mrad (dashed line).

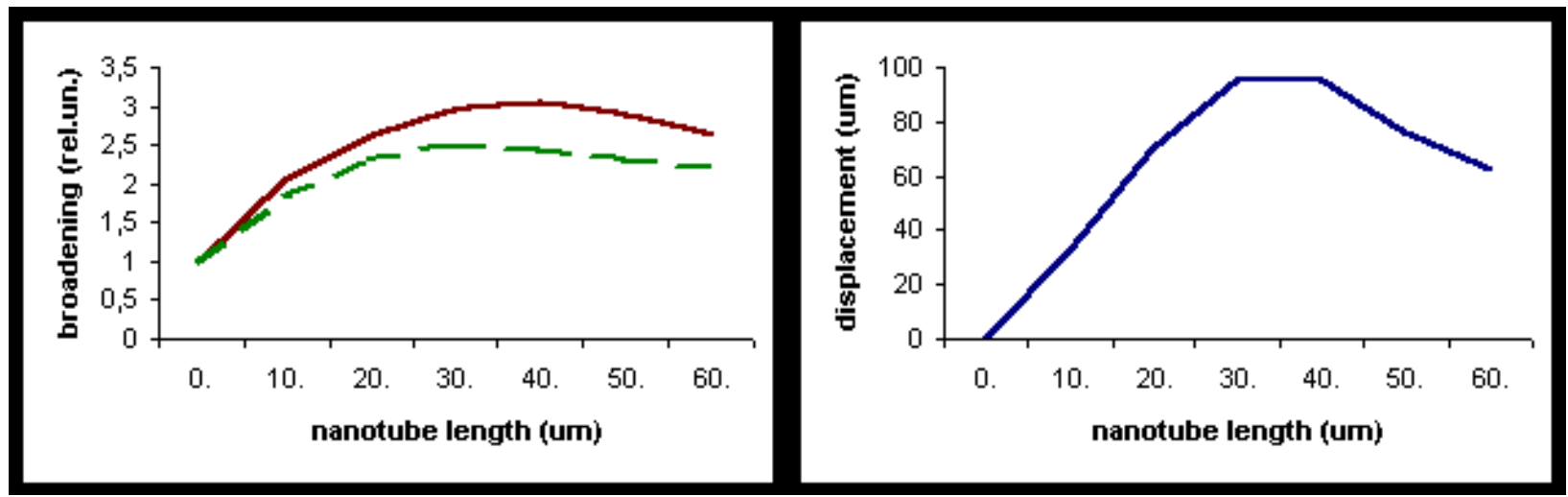

**Figure 4** Left figure: the rms size of the distribution for aligned sample, normalised to the same quantity for misaligned sample (top line, solid), and the same for the sample aligned at 0.04 mrad to the beam (bottom line). Right figure: the displacement (•m) of the distribution mean value for the sample aligned at 0.04 mrad to the beam, as a function of the nanotube length (•m).

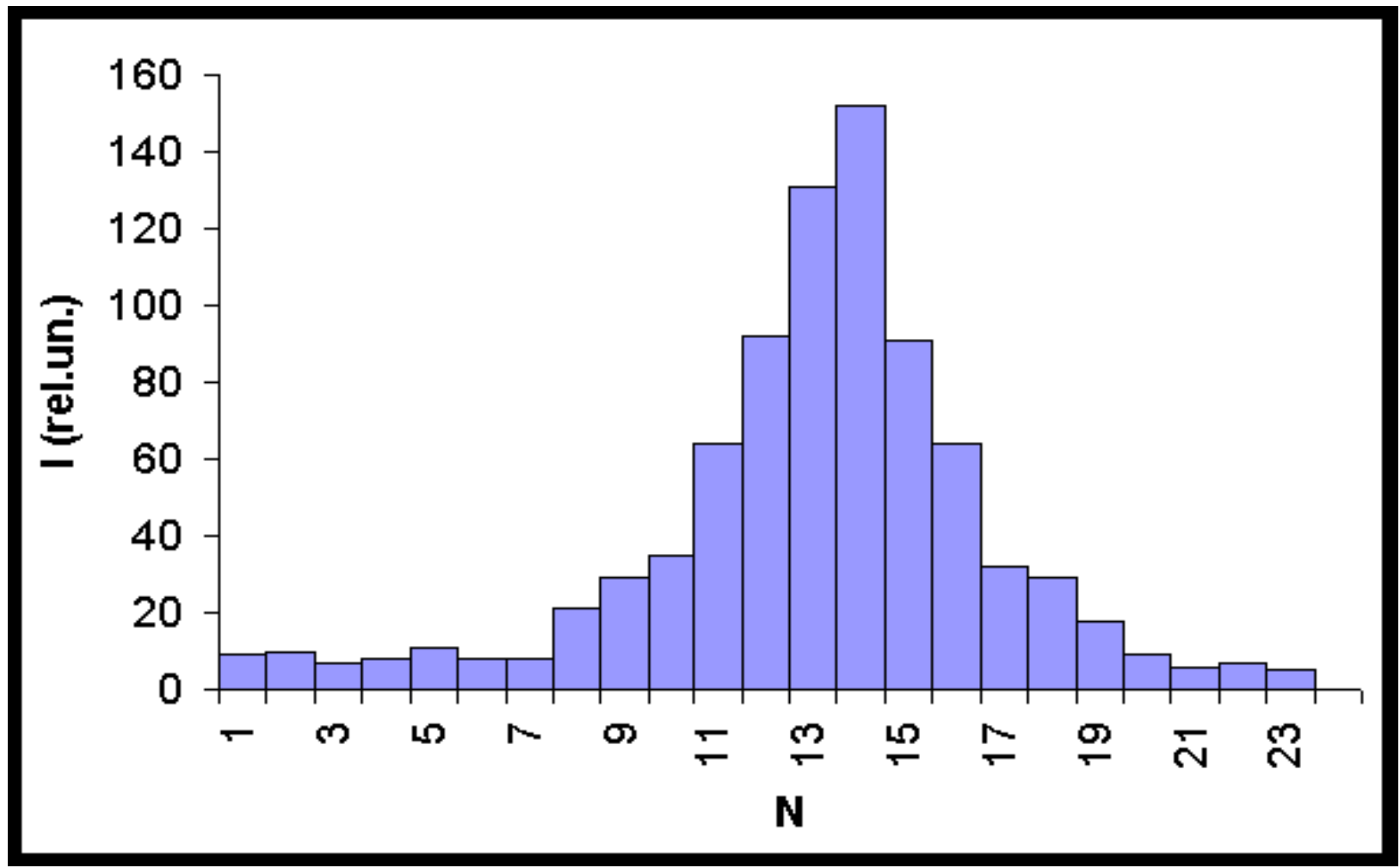

**Figure 5** Simulated distribution of 1.3-GeV protons at the detector downstream of a bent nanotube.

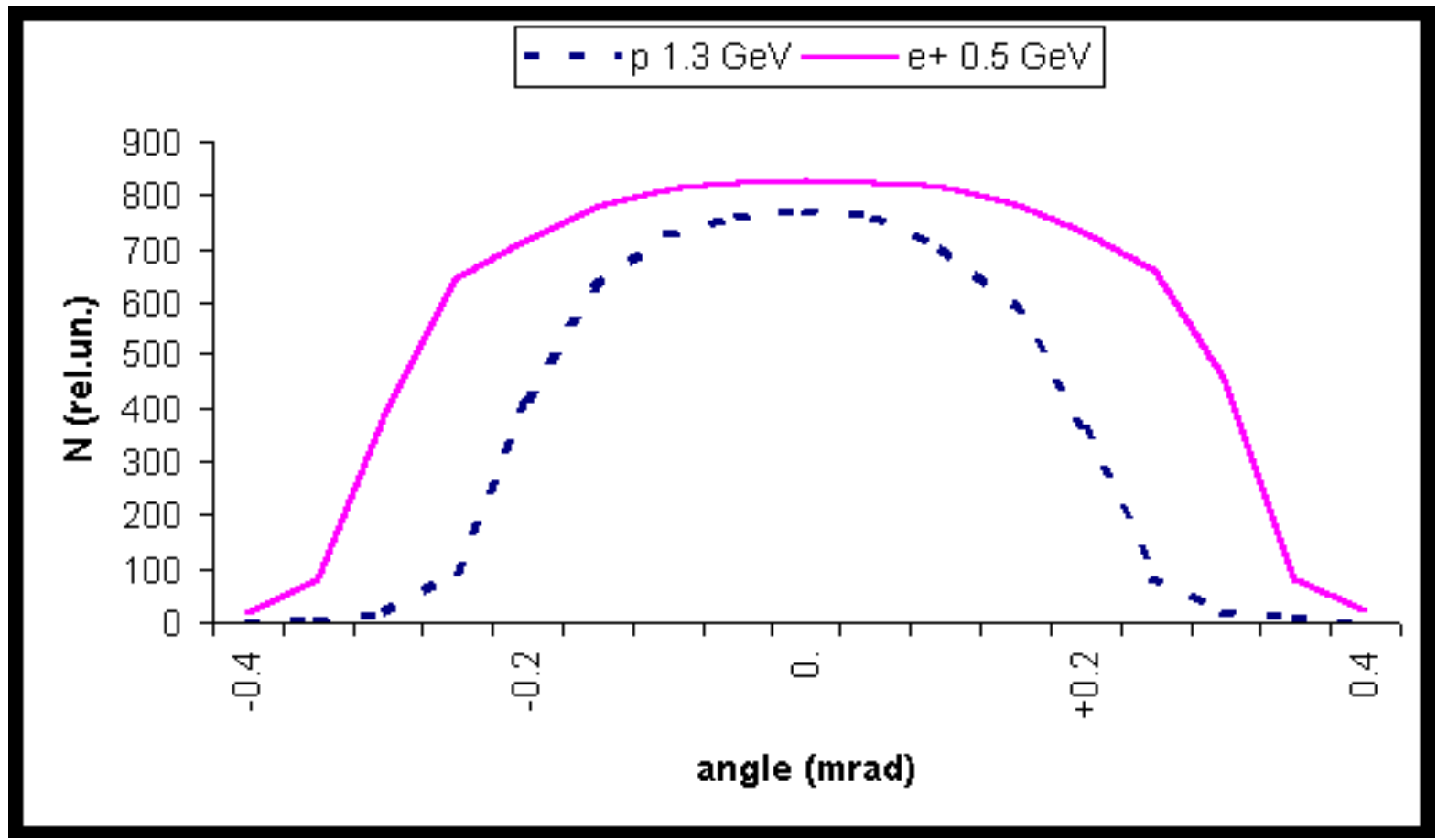

**Figure 6** The dependence of the channeled beam intensity on the orientation angle of the nanotube with respect to the beam of 0.5 GeV positrons (top, solid) and 1.3 GeV protons (bottom, dashed).